\documentclass[a4paper,11pt]{article}
\usepackage{graphicx,color,epsf}
\usepackage{jheppub} % for details on the use of the package, please
\usepackage[T1]{fontenc} % if needed

\allowdisplaybreaks

\title{\boldmath Hot scalar radiation around a cosmic string \\
setting bounds on the coupling parameter $\xi$}

\author{E. S. Moreira Jr.}

\affiliation{Instituto de Matem\'{a}tica e Computa\c{c}\~{a}o, 
Universidade Federal de Itajub\'a, 
Itajub\'a, MG 37500-903, Brazil}

\emailAdd{moreira@unifei.edu.br}

\abstract{ 
In this work,
by investigating the low temperature behaviour of a scalar field around
a cosmic string, and assuming stable thermodynamic equilibrium, it is shown that the coupling parameter of the field with curvature must be restricted to a certain range of values whose bounds are
determined by the deficit angle of the associated conical geometry.
}


\arxivnumber{1604.08089}

\begin{document} 
\maketitle
\flushbottom

\section{Introduction}
\label{int}
%it is desirable and sometimes necessary to
A recent publication regarding the low temperature regime of a scalar field near a reflecting wall in flat space has shown that the values of the  curvature coupling parameter $\xi$ must be restricted to a certain range when stable thermodynamic equilibrium is required \cite{del15}.
%(It should be mentioned that this is the first time that first principles %restrict the values of $\xi$
%as a thoroughly examination of the literature seemed to have shown.)
Perhaps the next step to investigate further this effect  would be to examine a background with genuine curvature.
%less trivial background, preferably one with genuine curvature.
A natural candidate is the background of a cosmic string
which is locally flat but globally curved as is well known (see, e.g., refs. 
\cite{smi90,vil94}).

The present work considers the thermal behaviour of the scalar radiation
in the conical spacetime of an infinite straight cosmic string. 
%%%COSMIC STRINGS
A cosmic string is a vortex-like object that
occurs as a topological defect in an Abelian $U$(1) scalar-gauge field model. 
At early stages of the universe, spontaneous symmetry breaking of the scalar
field led to topological defects, i.e., confined regions of false vacuum, among them cosmic strings.
The first general solution of
a self-gravitating general relativistic cosmic string has been found in ref. \cite{gar85}. If cosmic strings appeared at the early universe
they might have served as seeds for galaxies formation. However,
recent satellite observations have pointed out inconsistencies with the power spectrum of the CMB. Additionally, it is not clear if such topological defects could have survived since their appearance at the early stages of the universe. 
As a result interest in the subject diminished considerably.
Recently, this interest has grown with the finding that  
fundamental strings may play the role
of cosmic strings in the context of string theory or M-theory. Supersymmetric GUTs may even require the existence of cosmic strings
\cite{kib04,sak06}. It should be mentioned that cosmic strings in brane-world scenarios present new features which are not shared with their four-dimensional counterpart \cite{sla16}.
Apart from these considerations, cosmic strings are interesting objects in their own right and can be related to unusual physics,
such as the (2+1)-dimensional gravitating point particles or ``cosmons'' \cite{sta63,tho92}.
They may also play a role in a model for locally finite quantum gravity \cite{tho08}. It should be noted that the possibility 
that cosmic strings could create closed timelike curves has been ruled out in ref. \cite{des92}. 
%%%%%%%%%%%%%%%%%%%%%%%%%%%%%%%%%%%%%%%%%%%%%%%%%%%%%%%%%%%%%%%%%%%%%%%%%%%%%%%%%

In the next section the conical geometry of the  spacetime around a cosmic string is presented.
In section \ref{semt},
using formulas in the literature for the expectation value of the stress-energy-momentum tensor of a scalar field, 
$\left<{\cal T}^\alpha{}_\beta\right>$ \cite{fro95,gui95},
the low temperature regime of $\left<{\cal T}^\alpha{}_\beta\right>$
will be obtained. 
Then, in section \ref{bounds}, by requiring stable thermodynamic equilibrium,
it will result that not all values of the curvature coupling parameter $\xi$ are physically acceptable  if the conical singularity is present.
The way the bounds on $\xi$ vary with the corresponding deficit angle will be determined.
Section \ref{comments} closes the work with final remarks and addressing some speculations. 
(Throughout the text, $k_{B}=\hbar=c=1$.)

%\newpage

\section{The background}
\label{background}
The geometry around an infinite straight cosmic string is given by   \cite{smi90,vil94},
\begin{equation}
ds^{2}=dt^{2}-d\rho^{2}-\rho^{2}d\varphi^{2}-dz^{2},\hspace{1.7cm} \varphi \sim \varphi+2\pi \nu^{-1},
\label{ocone}
\end{equation}
where $\nu$ is related to the mass density $\mu$ of the string 
according to  $1/\nu=(1-4G\mu)$
and the spacial coordinates are the usual cylindrical coordinates. 
If $\mu\neq 0$ (i.e., $\nu \neq 1$)
the identification in eq. (\ref{ocone})
hides a conical singularity at $\rho=0$ (where the cosmic string lies) corresponding to a deficit angle, 
%${\cal D}$, 
\begin{equation}
{\cal D}=
2\pi(1-\nu^{-1})=8\pi G\mu .
\label{dangle}
\end{equation}
%$2\pi(1-1/\nu)=8\pi G\mu$.
Particles travelling around the cosmic string will be attracted 
by it when $\mu>0$ (attractive gravity); and scattered away from it when $\mu<0$ (repulsive gravity).
The corresponding values of $\nu$ are $1<\nu< \infty$ 
and  $0< \nu< 1$, respectively. When $\mu=0$ (i.e., $\nu=1$) the cosmic string is simply
absent and eq. (\ref{ocone}) corresponds to
Minkowski spacetime.

Although the conical spacetime is (globally) curved, a
(singular) Cartesian frame $(\bar{t},\bar{x}, \bar{y},\bar{z})$
is available where the ``flat'' coordinates are related to those in
eq. (\ref{ocone}) by 
\begin{equation}
\bar{t}=t, 
\hspace{0.5cm}
\bar{x}=\rho \cos\varphi, 
\hspace{0.5cm}
\bar{y}=\rho\sin\varphi, 
\hspace{0.5cm}
\bar{z}=z.
\hspace{0.5cm}
\label{ctransformation}
\end{equation}
%$\bar{t}=t$, $\bar{x}=\rho \cos\varphi$, $\bar{y}=\rho\sin\varphi$ and %$\bar{z}=z$. 
When $\nu>1$,  
from the point of view of the 
Cartesian frame the conical spacetime appears to be the Minkowski spacetime; but
with a missing wedge (or an extra wedge, when $\nu<1$).  
Note that the angle of the wedge is ${\cal D}$ in eq. (\ref{dangle}).
% in eq. (\ref{dangle}).

It should be pointed out that according to the physics of cosmic strings \cite{smi90,vil94},
\begin{equation}
G\mu \sim 10^{-6},
\label{density}
\end{equation}
thus $\nu$ is expected to be greater than unity but very close to it.

\section{Low temperature $\left<{\cal T}^\alpha{}_\beta\right>$}
\label{semt}

Vacuum polarization around conical singularities has been investigated since  the 1970s \cite{dow77,dow78}. The expression for the vacuum 
expectation value of the stress-energy-momentum tensor
%$\left<{\cal T}^\alpha{}_\beta\right>$ 
of a massless scalar field around a cosmic string has been found
to be \cite{smi90,hel86,lin87,fro87,dow87},
\begin{equation}
\left<{\cal T}^\alpha{}_\beta\right>_{o}=
\frac{1}{1440\pi^2 \rho^4}
\left[ (1-\nu^4)\ {\rm diag}\left(1,1,-3,1\right)+
10(6\xi-1)(1-\nu^2)\ {\rm diag}\left(2,-1,3,2\right)
\right],
\label{dowker}
\end{equation}
where coordinates were labelled as they appear in
the line element in eq. (\ref{ocone}), and where $\xi$ is the
curvature coupling parameter. It should be recalled that 
$\xi=0$ corresponds to the minimal coupling and 
$\xi=1/6$ to the conformal coupling (in which case $\left<{\cal T}^\alpha{}_\beta\right>_{o}$ is traceless as can be quickly checked).
When $\nu=1$, eq. (\ref{dowker}) vanishes showing that renormalization
has been implemented by removing the Minkowski vacuum [cf. eq. (\ref{ocone})]. It should be noted that backreaction 
in the metric tensor of eq. (\ref{ocone})
due to the vacuum polarization in eq. (\ref{dowker})
was considered in ref. \cite{his87}.

%References \cite{dav88,lin92} are
%early works on the behavior of the hot scalar radiation %around a cosmic string.
Generalizations of eq. (\ref{dowker}) to include finite temperature effects
around the cosmic string have been addressed in refs. \cite{fro95,gui95,dav88,lin92}. 
%According to ref. \cite{fro95}
The ensemble average  $\left<{\cal T}^\alpha{}_\beta\right>$ at temperature $T$ can be written as,
\begin{equation}
\left<{\cal T}^\alpha{}_\beta\right>=\left<{\cal T}^\alpha{}_\beta\right>_{o}+
\left<{\cal T}^\alpha{}_\beta\right>_{T},
%+\left<{\cal T}^\mu{}_\nu\right>_{{\tt thermal}},
\label{average}
\end{equation}
where the first term is just that in eq. (\ref{dowker}) and the second term vanishes when $T=0$.
Expressions for 
$\left<{\cal T}^\alpha{}_\beta\right>_{T}$
were given in ref. \cite{fro95} for arbitrary $\nu$, and in 
ref. \cite{gui95} when $\nu<2$.
\footnote{It should be noted that the calculations in
ref. \cite{gui95} agree with those in ref. \cite{fro95};
there is though a typo in eq. (29) of ref. \cite{gui95}, namely,
the factor $2\xi$ should be replaced by $4\xi$.}\ 
As stated by the authors of these references themselves such expressions are rather complicated; 
thus in other to simplify calculations this paper will consider only $0<\nu<2$, 
which  nevertheless covers the values of astrophysical interest 
[see eq. (\ref{density}) and the comment following it] and also addresses sharp cones 
[when $\nu\rightarrow 0$, ${\cal D} \rightarrow -\infty$, and
when $\nu=2$,  ${\cal D}=\pi$, cf. eq. (\ref{dangle})].

%$\left<T^\mu{}_\nu\right>_{{\tt thermal}}$ 
%is the familiar 
%stress-energy-momentum tensor of the blackbody radiation, i.e.,

According to ref. \cite{fro95},  $\left<{\cal T}^\alpha{}_\beta\right>$ 
%in eq. (\ref{average})
is a diagonal matrix whose diagonal components are given by,
\begin{eqnarray}
&&
\left<{\cal T}^t{}_t\right>=
\frac{\pi^2}{30}T^4+
2\pi T^{4}\left[I_{2}-(1-4\xi)(I_{3}+I_{4})\right]\nu\sin \nu\pi,
%\label{edensity0}
\nonumber 
\\&&
\left<{\cal T}^\rho{}_\rho\right>=
-\frac{\pi^2}{90}T^4+
2\pi T^{4}\left[I_{1}-4\xi I_{3}\right]\nu\sin \nu\pi,
%&&
%\label{rpressure0}
\nonumber 
\\&&
\left<{\cal T}^\varphi{}_\varphi\right>=
-\frac{\pi^2}{90}T^4+
2\pi T^{4}\left[-2I_{1}-I_{2}+4\xi (2I_{3}+I_{4})\right]\nu\sin \nu\pi,
%\label{llpressure0}
\nonumber 
\\
&&
\left<{\cal T}^z{}_z\right>=
-\frac{\pi^2}{90}T^4+2\pi T^{4}\left[I_{1}
-(1-4\xi)(I_{3}+I_{4})
%-I_{3}-I_{4}+4\xi (I_{3}+I_{4})
\right]\nu\sin \nu\pi,
\label{fpz}
\end{eqnarray}
where
\begin{eqnarray}
&&I_{1}:=
\int_{0}^{\infty} du\
\frac{w+\sinh w}{w^3(\cosh w -1)}\
\frac{1}{\cosh \nu u - \cos \nu \pi},
\nonumber\\
&&
I_{2}:=
\int_{0}^{\infty} du\
\frac{\sinh w}{w(\cosh w -1)^{2}}\
\frac{1}{\cosh \nu u - \cos \nu \pi},
\nonumber\\
&&
I_{3}:=
\int_{0}^{\infty} du\
\frac{w+\sinh w}{w^3(\cosh w -1)}\
\frac{\cosh^2(u/2)}{\cosh \nu u - \cos \nu \pi},
\nonumber\\
&&
%\hspace{0.1cm}
I_{4}:=
\int_{0}^{\infty} du\
\frac{\sinh w}{w(\cosh w -1)^{2}}\
\frac{\cosh^2(u/2)}{\cosh \nu u - \cos \nu \pi},
%&&
\label{I}
\end{eqnarray}
and
\begin{equation}
w:=4\pi T\rho\cosh(u/2).
\label{w}
\end{equation}
%$w:=4\pi T\rho\ \cosh(u/2)$.
It is worth remarking that $I_{i}$ ($i=1,2,3,4$) depends on
$T$, $\rho$ and $\nu$.

For the purpose of this work, only the low temperature behaviour
(more precisely,  when $T\rho\ll 1$) of $\left<{\cal T}^\alpha{}_\beta\right>$ 
%in eq. (\ref{average}) 
will be needed. Such a regime is obtained by realizing that, 
when $T\rho\rightarrow 0$, each $I_{i}$ defined in eq. (\ref{I})
can be very well approximated as
$V_{i}+c_{i}$ where $V_{i}\propto 1/(T\rho)^4$ whereas 
$c_{i}$ does not depend either on $T$ or $\rho$.
Indeed, by expanding the integrand of
$I_{1}$ in powers of 
$w$ keeping only leading terms as $w\rightarrow 0$
[i.e., as $T\rho\rightarrow 0$, cf. eq. (\ref{w})] and then 
by integrating over $u$ each term in the expansion,
one finds approximately that,
\begin{equation}
I_{1}= V_{1}+c_{1},
\label{I1} 
\end{equation}
with
\begin{equation}
V_{1}=\frac{1}{64(\pi T\rho)^4}
\int_{0}^{\infty} du\
%\frac{1}{\cosh^4 \frac{u}{2}}\
\frac{\cosh^{-4}(u/2)}{\cosh \nu u - \cos \nu \pi},
 \label{V1} 
\end{equation}
and
\begin{equation}
c_1(\nu)=
\frac{1}{180}
\int_{0}^{\infty} du\
%\frac{1}{\cosh^4 \frac{u}{2}}\
\frac{1}{\cosh \nu u - \cos \nu \pi}=
\frac{\left(\nu^{-1} -1\right)\pi}{180\sin\nu\pi},
\label{c1}
\end{equation}
where the last equality in eq. (\ref{c1}) has been obtained by using 
ref. \cite{gra07} \footnote{A closed form for each $V_{i}$ is 
also available; but it will not be explicitly needed here.}.
An examination  of eq. (\ref{c1}) yields,
%For future use in this paper note that,
\begin{equation}
c_1(1)=\frac{1}{180}.
\label{c11}
\end{equation}
Likewise, expanding now the integrand of $I_{2}$ in 
the second expression of
eq. (\ref{I}) in powers of $w$, 
%following the same steps
%that led to eq. (\ref{I1}), 
it follows that $I_{2}= V_{2}+c_{2}$ where 
$V_{2}=V_{1}$ and $c_{2}=-3c_{1}$, i.e.,
\begin{equation}
I_{2}= V_{1}-3c_{1},
\label{I2} 
\end{equation}
as $T\rho\rightarrow 0$.

Repeating the power expansion procedure with 
$I_{3}$ in 
the third expression of
eq. (\ref{I}), when $T\rho\rightarrow 0$, it results that,
\begin{equation}
I_{3}= V_{3}+c_{3},
\label{I3} 
\end{equation}
where,
\begin{equation}
V_{3}=\frac{1}{64(\pi T\rho)^4}
\int_{0}^{\infty} du\
%\frac{1}{\cosh^4 \frac{u}{2}}\
\frac{\cosh^{-2}(u/2)}{\cosh \nu u - \cos \nu \pi},
 \label{V3} 
\end{equation}
and
\begin{equation}
c_3(\nu)=
\frac{1}{180}
\int_{0}^{\infty} du\
%\frac{1}{\cosh^4 \frac{u}{2}}\
\frac{\cosh^2(u/2)}{\cosh \nu u - \cos \nu \pi}=
\frac{-\pi}{360\sin\nu\pi}. 
\label{c3}
\end{equation}
Again ref. \cite{gra07} has been used to obtain the last equality in eq. (\ref{c3})
\footnote{Incidentally a typo in formula 3.514-2 of ref. \cite{gra07}
is reported: $\pi t_{2}$ should be replaced by $\pi-t_{2}$.}. 

An inspection of eq. (\ref{c3}) shows clearly that it diverges as $\nu\rightarrow 1$.
This issue can be circumvented by writing, 
\begin{equation}
c_3(\nu):=\lim_{T\rho\ \rightarrow\  0}
\left(I_3-V_3\right).
\label{c}
\end{equation}
Considering now eq. (\ref{c}) without expanding the integrand of $I_3$ in 
the third expression of 
eq. (\ref{I}) and noting eq. (\ref{V3}),
a numerical analysis shows that, 
%approximately,
\begin{equation}
c_3(1)\approx\frac{1}{25}.
\label{c33}
\end{equation}
Indeed, eq. (\ref{c3}) reproduces  $c_3$ satisfactorily only when $\nu>1$ 
and $\nu$ is not very close to unity. 
[For instance, when $\nu=1.1$, $c_3\approx 0.022$
whereas eq. (\ref{c3}) yields approximately 0.028, improving quickly as
$\nu\rightarrow 2$.]
It should be pointed out that the behaviour of $c_{3}$ in eq. (\ref{c})
over the range of interest, 
$0<\nu<2$, resembles very much that of $c_{1}$ in eq. (\ref{c1}).

Addressing now $I_{4}$ in 
the fourth expression of
eq. (\ref{I}) and proceeding along the same lines as above one finds that
$I_{4}= V_{4}+c_{4}$, where 
$V_{4}=V_{3}$ and $c_{4}=-3c_{3}$, i.e.,
\begin{equation}
I_{4}= V_{3}-3c_{3},
\label{I4} 
\end{equation}
which holds as $T\rho\rightarrow 0$.

The low temperature regime of the components in eq. (\ref{fpz}) can now be
determined by considering $I_{i}$ in eqs. (\ref{I1}), (\ref{I2}), (\ref{I3}) and (\ref{I4}),
resulting in eq. (\ref{average}), where 
$\left<{\cal T}^\alpha{}_\beta\right>_{o}$
arises from the terms involving 
$V_{i}$,
\footnote{Using the closed form for $V_{i}$ (see, e.g., ref. {\cite{gui95}}) eq. (\ref{dowker})
is obtained.} and the nonvanishing components of 
$\left<{\cal T}^\alpha{}_\beta\right>_{T}$
%can be determined  by manipulating formulas in ref. \cite{fro95}, 
%leading to a diagonal matrix whose diagonal components 
are given by,
\begin{eqnarray}
&&
\left<{\cal T}^t{}_t\right>_{T}=
\frac{\pi^2}{30}T^4
\left[1+\frac{60}{\pi}\left[-3c_1+2c_3(1-4\xi)\right]\nu\sin \nu\pi\right],
\nonumber
%\label{edensity}
\\&&
\left<{\cal T}^\rho{}_\rho\right>_{T}=
\left<{\cal T}^\varphi{}_\varphi\right>_{T}=
-\frac{\pi^2}{90}T^4
\left[1-\frac{180}{\pi}\left[c_1-4\xi c_3\right]\nu\sin \nu\pi\right],
%&&
%\label{rpressure}
\nonumber
\\
&&
\left<{\cal T}^z{}_z\right>_{T}=
-\frac{\pi^2}{90}T^4
\left[1-\frac{180}{\pi}\left[c_1+2c_3(1-4\xi)\right]\nu\sin \nu\pi\right].
\label{lpressure}
\end{eqnarray}

As expected [cf. eq. (\ref{ocone})],
%considering the discussion above regarding $c_2$ and $c_3$, 
a glance on
eqs. (\ref{dowker}), (\ref{average}) and (\ref{lpressure}) shows that,
when $\nu=1$,  $\left<{\cal T}^\alpha{}_\beta\right>_{T}$
and $\left<{\cal T}^\alpha{}_\beta\right>$
simply reduce to the familiar stress-energy-momentum tensor 
of blackbody radiation, i.e.,
\begin{equation}
\left<{\cal T}^\alpha{}_\beta\right>=
\left<{\cal T}^\alpha{}_\beta\right>_{T}=
\frac{\pi^2}{90}T^4 {\rm diag}(3,-1,-1,-1),
\hspace{1.0cm}
\nu=1.
\label{blackbody}
\end{equation}

%For future use in this paper

\section{Bounds on $\xi$}
\label{bounds}
%Although the conical spacetime is (globally) curved, a
%(singular) Cartesian frame $(\bar{t},\bar{x}, \bar{y},\bar{z})$
%is available where the ``flat'' coordinates are related to those in
%eq. (\ref{ocone}) by $\bar{t}=t$, $\bar{x}=\rho \cos\varphi$, %$\bar{y}=\rho\sin\varphi$ and $\bar{z}=z$. When $\nu>1$,  from the point of %view of the 
%Cartesian frame the conical spacetime appears to be the Minkowski spacetime; %but
%with a missing wedge (or an extra wedge, when $\nu<1$).  
%Note that the angle of the wedge is ${\cal D}$ in eq. (\ref{dangle}).
This section begins with a short digression on  global conservation
of energy and momentum of a physical system in a locally flat background. By using  flat coordinates,
the local conservation of energy and momentum is given by the familiar expression,
\begin{equation}
{\cal T}^\alpha{}^\beta,_{\beta}=0.
\label{emconservation}
\end{equation}
The energy flux density is defined as
${\bf S}=({\cal T}^0{}^1,{\cal T}^0{}^2,{\cal T}^0{}^3)$, and the
momentum ${\bf P} $ in a volume $V$ limited by a surface $\sigma$ is,
\begin{equation}
{\bf P} =\int_{V}{\bf S}\ d^{3}\hspace{-0.05cm}x.
\label{momentum}
\end{equation}
Global conservation of energy is obtained from Eq.~(\ref{emconservation}), i.e.,
\begin{equation}
\frac{d}{dt}\int_{V}{\cal T}^0{}^0 d^{3}\hspace{-0.05cm}x
=-\oint_{\sigma}{\cal T}^0{}^i n^{i} da
=-\oint_{\sigma}{\bf S}\cdot{\bf n}\ da,
\label{econservation}
\end{equation}
where ${\bf n}$ is the outward normal to the surface $\sigma$.
Defining,
\begin{equation}
{\bf F}=\frac{d{\bf P}}{dt},
\label{force}
\end{equation}
global conservation of momentum also follows from
Eq.~(\ref{emconservation}), namely,
\begin{equation}
{\rm F}^{i}=-\oint_{\sigma}{\cal T}^i{}^j n^{j} da.
\label{mconservation}
\end{equation}
%where, as in Eq.~(\ref{econservation}), $i$ and $j$ run from 1 to 3,
%and repeated indices indicates summation.

In order to use these formulas in the present context, 
%the arguments in section 5 of ref. \cite{del15} 
it is convenient to work with 
\begin{equation}
\left<\bar{{\cal T}}^\alpha{}^\beta\right>=\left<\bar{{\cal T}}^\alpha{}^\beta\right>_{o}+
\left<\bar{{\cal T}}^\alpha{}^\beta\right>_{T},
%+\left<\bar{{\cal T}}^\mu{}^\nu\right>_{{\tt thermal}},
\label{average2}
\end{equation}
%$\left<\bar{{\cal T}}^\alpha{}^\beta\right>$,
which can be obtained from eq. (\ref{average}) by 
transforming to the Cartesian coordinates
[cf. eq. (\ref{ctransformation})]. 
The vacuum contribution in eq. (\ref{average2}) 
will not be needed in the arguments below 
%(since it does not depend on $T$) 
and the thermal contribution has a diagonal form, such that,
\begin{eqnarray}
\left<{\bar{\cal T}}^{\bar{t}}{}^{\bar{t}}\right>_{T}=
\left<{\cal T}^t{}_t\right>_{T},
\hspace{0.8cm}
\left<{\bar{\cal T}}^{\bar{x}}{}^{\bar{x}}\right>_{T}=
\left<{\bar{\cal T}}^{\bar{y}}{}^{\bar{y}}\right>_{T}=
-\left<{\cal T}^\rho{}_\rho\right>_{T},
\hspace{0.8cm}
\left<{\bar{\cal T}}^{\bar{z}}{}^{\bar{z}}\right>_{T}=
-\left<{\cal T}^z{}_z\right>_{T},
&&
\label{newstress}
\end{eqnarray}
where the components grouped in eq. (\ref{lpressure}) should be observed.
For instance, it follows that the rate of change of the energy density
$\left<{\bar{\cal T}}^{\bar{t}}{}^{\bar{t}}\right>$ with $T$ is given by,
\begin{equation}
c_{V}=
\frac{2\pi^2}{15}T^3
\left[1+\frac{60}{\pi}\left[-3c_1+2c_3(1-4\xi)\right]\nu\sin \nu\pi\right].
\label{heat}
\end{equation}
Perhaps it should be remarked that, when $\nu=1$, eq. (\ref{heat}) becomes
the Planckian specific heat per unit volume [cf. eq. (\ref{blackbody})].

As in ref. \cite{del15},  
one considers a small cubic region of the scalar radiation
with sides facing the Cartesian planes and assumes that the temperature
$T_{in}$ inside the cube differs slightly from $T_{out}$, which is the temperature outside.
The sides of the cube have area $A$ and each of the sides is imagined to be contained in a 
very thin rectangular
parallelepiped with volume $A\delta$, where $\delta$ is very small. 
%Thus, there are six of these parallelepipeds. 
The three sides of the cube whose outward normal coincides with  the
usual Cartesian coordinate vectors ${\bf i}$, 
${\bf j}$ and ${\bf k}$
%as well as their corresponding parallelepipeds, 
will be labelled  below (1), (2) and (3), respectively.
It follows from 
eq. (\ref{mconservation}) that the rate of change
of momentum with time in each corresponding parallelepiped is, 
\begin{eqnarray}
&&{\bf F}_{(1)}=\left[\left<{\bar{\cal T}}^{\bar{x}}{}^{\bar{x}}\right>_{T_{in}}
-\left<{\bar{\cal T}}^{\bar{x}}{}^{\bar{x}}\right>_{T_{out}}\right]\hspace{-0.1cm}A
%\hspace{-0.1cm}A 
{\bf i},
\hspace{0.9cm}
{\bf F}_{(2)}=
\left[\left<{\bar{\cal T}}^{\bar{y}}{}^{\bar{y}}\right>_{T_{in}}
-\left<{\bar{\cal T}}^{\bar{y}}{}^{\bar{y}}\right>_{T_{out}}\right]\hspace{-0.1cm}A
{\bf j},
\nonumber
\\
&&
\hspace{3.1cm}
{\bf F}_{(3)}=
\left[\left<{\bar{\cal T}}^{\bar{z}}{}^{\bar{z}}\right>_{T_{in}}
-\left<{\bar{\cal T}}^{\bar{z}}{}^{\bar{z}}\right>_{T_{out}}\right]\hspace{-0.1cm}A
{\bf k}.
\label{parallelepipeds}
\end{eqnarray}
%where $N=4$ has been set in 
%Eqs.~(\ref{pep}), (\ref{edw}) and (\ref{es}).
If at some instant ${\bf S}$ vanishes,
%inside the cubic region, 
after a short interval of time $\Delta t$ ($i=1,2,3$),
\begin{equation}
{\bf S}_{(i)}=\frac{1}{A\delta} {\bf F}_{(i)} \Delta t,
\label{energyflux}
\end{equation}
where Eqs.~(\ref{momentum}) and (\ref{force}) have been used.
Then, by using 
eqs. (\ref{parallelepipeds}) and (\ref{energyflux}), 
one finds
%the energy flux per unit
%of time out of the cube, $\oint_{S}{\bf S}\cdot{\bf n}\ da$, is 
%given by 
%\begin{equation}
%\Phi= \left[1-2(1-4\xi)\right]\left(T^{4}_{in}-T^{4}_{out}\right),
%\label{flux}
%\end{equation}
%bigger than that outside, i.e.,
%$T_{in}>T_{out}$. 
that the  power radiated out of the cube,
$\oint_{Sides}{\bf S}\cdot{\bf n}\ da$, is proportional to
the algebraic sum of the differences of pressures inside and outside, namely,
\begin{equation}
\Phi=\left[\left<{\bar{\cal T}}^{\bar{x}}{}^{\bar{x}}\right>_{T_{in}}
-\left<{\bar{\cal T}}^{\bar{x}}{}^{\bar{x}}\right>_{T_{out}}\right]
+
\left[\left<{\bar{\cal T}}^{\bar{y}}{}^{\bar{y}}\right>_{T_{in}}
-\left<{\bar{\cal T}}^{\bar{y}}{}^{\bar{y}}\right>_{T_{out}}\right]
+
\left[\left<{\bar{\cal T}}^{\bar{z}}{}^{\bar{z}}\right>_{T_{in}}
-\left<{\bar{\cal T}}^{\bar{z}}{}^{\bar{z}}\right>_{T_{out}}\right],
\label{balance}
\end{equation}
up to a positive overall factor. 
%\cite{del15}.
Taking into account eq. (\ref{newstress}) in eq. (\ref{balance}) and dropping again a positive overall factor, it follows that,
\begin{equation}
\Phi=
\left[1-\frac{60}{\pi}\left[3c_1+2c_3(1-8\xi)\right]\nu\sin \nu\pi\right]
\left(T_{in}^4-T_{out}^4\right).
\label{flux}
\end{equation} 
%up to an overall positive factor.

Now say that, for example,
$T_{in}>T_{out}$.
%\begin{equation}
%T_{in}>T_{out}.
%\label{temperatures}
%\end{equation}
If $c_V>0$, eq. (\ref{heat}) yields,
\begin{equation}
\xi\sin \nu\pi<\frac{\pi}{480 c_3\nu}+
\frac{2c_3-3c_1}{8c_3} \sin \nu\pi.
\label{inequality1}
\end{equation}
As by hypothesis $c_V>0$, then $\Phi$ in eq. (\ref{flux}) must be positive
such that energy leaves the cube 
[cf. eq. (\ref{econservation})]
ensuring that $T_{in}$ decreases and thermodynamic equilibrium is eventually restored, i.e.,
\begin{equation}
\xi\sin \nu\pi>\frac{-\pi}{960 c_3\nu}+
\frac{3c_1+2c_3}{16c_3} \sin \nu\pi.
\label{inequality2}
\end{equation}
The values of $\xi$ consistent with stable thermodynamic equilibrium are determined
by requiring that eqs. (\ref{inequality1}) and (\ref{inequality2}) are simultaneously 
satisfied. [Proceeding along the same lines, by assuming $c_V<0$ instead,  
%in eq. (\ref{heat})
it results that the inequalities corresponding to stable thermodynamic equilibrium cannot  be simultaneously satisfied by any $\xi$.]

\subsection{Permissible values of $\xi$ when $1<\nu<2$}
\label{gravity}
\begin{figure}[h]
\center
\includegraphics[scale=1.0]{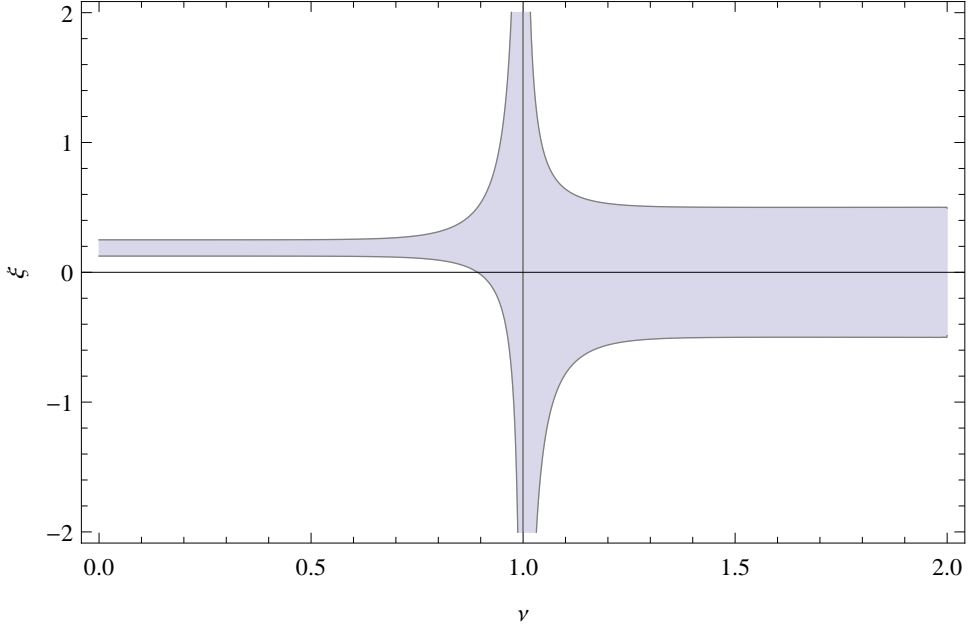}
%{graficofinal.eps}
%\includegraphics[scale=0.40]{grafico1.eps}
\caption{\footnotesize{Permissible values of $\xi$ for a given $\nu$} ---
gray region. It should be noted that the conformal
coupling, $\xi=1/6$, belongs to the whole permissible region; but that the minimal
coupling, $\xi=0$, does not.
}
\label{figure}
\end{figure}

When $1<\nu<2$, eqs. (\ref{inequality1}) and (\ref{inequality2})
lead to 
\begin{equation}
\frac{\pi}{480 c_3\nu\sin \nu\pi}+
\frac{2c_3-3c_1}{8c_3}<\xi<\frac{-\pi}{960 c_3\nu\sin \nu\pi}+
\frac{3c_1+2c_3}{16c_3},
\label{inequality3}
\end{equation}
whose bounds 
%in eq. (\ref{inequality3})
are plotted in figure \ref{figure},
on the right of the straight line $\nu=1$.
Considering eqs. (\ref{c1}) and (\ref{c3}) in eq. (\ref{inequality3}),
one arrives at,
\begin{equation}
-\frac{1}{2}<\xi<\frac{1}{2},
\hspace{2.0cm}
\nu\rightarrow 2.
\label{inequality31}
\end{equation}
Noting eqs. (\ref{c11}) and (\ref{c33}), it results from
eq. (\ref{inequality3}) approximately that,
\begin{equation}
-\frac{1}{19(\nu-1)}<\xi<\frac{1}{38(\nu-1)},
\hspace{2.0cm}
\nu\rightarrow 1,
\label{inequality32}
\end{equation}
showing the leading behaviour of the bounds just on the right of the
straight line $\nu=1$ in figure \ref{figure}.
It is worth remarking that, as $\nu$ increases, the transition from the range in 
eq. (\ref{inequality32}) to that in eq. (\ref{inequality31}) is
rather quick  (cf. figure \ref{figure}).

%%%%%%%%%%%%%%%FIGURE%%%%%%%%%%%%%%%%%%%%%%%
%%%%%%%%%%%%%%%%%%%%%%%%%%%%%%%%%%%%%%%%%%%%%%%%%%%

\subsection{Permissible values of $\xi$ when $0<\nu<1$}
\label{antigravity}
When $0<\nu<1$, now eqs. (\ref{inequality1}) and (\ref{inequality2})
lead to 
\begin{equation}
\frac{-\pi}{960 c_3\nu\sin \nu\pi}+
\frac{3c_1+2c_3}{16c_3}
<\xi<\frac{\pi}{480 c_3\nu\sin \nu\pi}+
\frac{2c_3-3c_1}{8c_3},
\label{inequality4}
\end{equation}
which is illustrated 
%in eq. (\ref{inequality3})
in figure \ref{figure},
on the left of the straight line $\nu=1$.
Instead of eqs. (\ref{inequality32}), it follows from eq. (\ref{inequality4}),
\begin{equation}
-\frac{1}{38(1-\nu)}<\xi<\frac{1}{19(1-\nu)},
\hspace{2.0cm}
\nu\rightarrow 1.
\label{inequality41}
\end{equation}
Although there is no closed form available for $c_3$ when $\nu <1$, knowing that
it behaves as $1/\nu ^2$ when $\nu\rightarrow 0$, and using  
eq. (\ref{c1}), one readily obtains from eq. (\ref{inequality4}) that,
\begin{equation}
\frac{1}{8}<\xi<\frac{1}{4},
\hspace{2.0cm}
\nu\rightarrow 0.
\label{inequality42}
\end{equation}
Again, as $\nu$ decreases from $\nu=1$, the onset of the range in eq. (\ref{inequality42}) is rather quick (cf. figure \ref{figure}).

\section{Conclusion}
\label{comments}
This paper shows that the  conical geometry around an infinite straight cosmic string 
restricts the values of the curvature coupling parameter $\xi$ when stable thermodynamic equilibrium 
of the scalar field with the background is required. 
The analysis is based on  the low temperature behaviour of
$\left<{\cal T}^\alpha{}_\beta\right>$ obtained from known formulas
in the literature. The study  adds new arguments to the 
conjecture that in any nontrivial background hot scalar radiation sets bounds on $\xi$ \cite{del15}.

The behaviour of the bounds on $\xi$, in the associated conical background 
[cf. eqs. (\ref{ocone}) and (\ref{dangle})],
is illustrated in figure \ref{figure} for $-\infty <{\cal D}<\pi$, i.e., $0<\nu<2$.
The highlights are the asymptotic regimes in eqs. (\ref{inequality31}) and
(\ref{inequality32}) (corresponding to attractive gravity), and those in 
eqs. (\ref{inequality41}) and
(\ref{inequality42}) (corresponding to repulsive gravity). It can be seen that
$\xi$ is restricted as long as the conical singularity is present,
i.e., $\nu\neq 1$. Nevertheless, for cosmic strings of physical interest [cf. eq. (\ref{density})],
only very large values of $|\xi|$ are unphysical.

Examining figure \ref{figure} one sees that $\nu$ is constrained even for the minimal coupling. That is,
for $\xi=0$ stable thermodynamic equilibrium requires
$\nu>\nu_{0}\approx 0.893$.  
This may appear unexpected since, when $\xi=0$,
${\cal T}^\alpha{}_\beta$ in the geometric background of a cosmic string reduces to the familiar canonical stress-energy-momentum tensor
in Minkowski spacetime. In order to clarify this fact it should be recalled that 
the presence of the cosmic string
``deforms'' the vacuum, resulting in a vacuum expectation value of the observable ${\cal T}^\alpha{}_\beta$ 
which is different from that corresponding to the Minkowski vacuum (in a manner that resembles the well known Casimir effect). 
Indeed, when $\nu\neq 1$ (i.e., when the cosmic string is present), the vacuum expectation value in eq. (\ref{dowker}) 
is nonzero when $\xi=0$. The cosmic string
also affects thermal averages when $\xi=0$ [cf. eq. (\ref{lpressure})] 
and eventually sets a constraint on $\nu$, as noted above. 
If one insists in having scalar radiation in stable thermodynamic equilibrium
around a cosmic string for which  $\nu<\nu_{0}$, then nonminimal coupling will be needed.

%In coming to the end of this paper
%NOTE: the cosmic string polarizes the spacetime around it. This vacuum polarization
%is not due to the coupling with the local curvature (although depends on  $\xi$);
%but due to the associated non trivial Christoffel symbols, in a manner that recalls the  Aharonov-Bohm effect.

It should be noticed a rather curious feature regarding the range in eq. (\ref{inequality42}). Namely,
it is precisely that corresponding to the allowed values of $\xi$ when the scalar
field is near a reflecting Dirichlet wall in flat space \cite{del15}. [This fact looks less surprising
if one recalls that $\nu\rightarrow 0$ corresponds to an infinite repulsive gravitational
force (cf. section \ref{background}).]

%In comming to the end of this paper
It is inevitable to conjecture that the range in eq.  (\ref{inequality31}) 
will still hold when $\nu>2$, though that needs confirmation by further calculations and analysis.

As seen in section \ref{bounds}, values of $\xi$ corresponding to negative $c_{V}$ must be excluded.
Such values are unphysical since they would lead eventually to inhomogeneities, spoiling thermodynamic
equilibrium. At this point one may recall that for a 
conformally coupled scalar field in the Hartle-Hawking thermal state 
around a Schwarzschild black hole, $c_{V}<0$ at the event horizon \cite{pag82}.
It must be noted, however, that the cubic region of scalar radiation described in section \ref{bounds}
was examined by a stationary observer, to whom the ground state in the black hole background is the so called 
Boulware state (which, at infinity, becomes the Minkowski vacuum) \cite{can80}. It seems then that one
should ``renormalize'' the energy density of the Hartle-Hawking state by subtracting
that corresponding to the Boulware state \cite{dow78}. In so doing, a positive $c_{V}$ arises instead.
A word of caution about this fact should be noted. Although the scalar radiation is in stable thermodynamic
equilibrium with itself; 
%(i.e., $c_{V}>0$); 
it is not in stable thermodynamic equilibrium with the 
black hole, whose heat capacity is negative, as is fairly well known.

%$\left<{\cal T}^t{}^t\right>$,

Although only a massless field was considered above,
the results also apply to a massive scalar field 
with mass $M\ll T$ (and possibly to a massive scalar field of arbitrary mass,
as happens in the case considered in ref. \cite{del15}).

%Instability is a warning that a sudden shift (deviation) may about to occur.
%Hawking radiation is imminent.
%Perhaps a speculation regarding black holes is in order.
%would hardly be observed

%In coming to the end of this paper
%NOTE: the cosmic string polarizes the spacetime around it. This vacuum polarization
%is not due to the coupling with the local curvature (although depends on  $\xi$);
%but due to the associated non trivial Christoffel symbols, in a manner that recalls the  Aharonov-Bohm effect.

%\newpage

\acknowledgments
Work partially supported by FAPEMIG.

%\vspace{1cm}
%\noindent{\bf Acknowledgements} -- Work partially supported by FAPEMIG.

%\newpage

%\end{multicols}

\end{document}